\documentclass[english, oneside]{article}   	
\usepackage{geometry}                	
\usepackage[parfill]{parskip}    		
\usepackage{graphicx}				
\usepackage{subfig}
\usepackage{bm}
\usepackage{float}
\usepackage{amsbsy}
\usepackage{amsmath}
\usepackage{commath}
\usepackage{esint}
\usepackage{xcolor}													
\usepackage{amssymb}

\newcommand{\bra}[1]{\langle #1|}
\newcommand{\ket}[1]{|#1\rangle}
\newcommand{\braket}[2]{\langle #1|#2\rangle}
\newcommand{\Schrs}{Schr\"odinger's }
\newcommand{\Schr}{Schr\"odinger }

\newcommand{\sg}{Stern-Gerlach }

\title{\center A method for measuring the real part of the weak value of spin using non-zero rest mass particles.}

\author{V. Monachello$^1$\footnote{E-mail address vincenzo.monachello.14@ucl.ac.uk}, R. Flack$^1$\footnote{E-mail address r.flack@ucl.ac.uk}, B. J. Hiley$^{1, 2}$\footnote{E-mail address b.hiley@bbk.ac.uk} \,and  R. E. Callaghan$^2$\footnote{E-mail address r.callaghan@physics.org}.\\\\ 1. Dept. of Physics and Astronomy, University College, Gower Street, \\London WC1E 6BT \\2. TPRU, Birkbeck, University of London, Malet Street, \\London WC1E 7HX \\ }
\date{January 2017} 							

\begin{document}

\maketitle

\begin{abstract}
A method for measuring the real part of the weak (local) value of spin is presented using a variant on the original \sg apparatus. The experiment utilises metastable helium in the $\rm 2^{3}S_{1}$ state. A full simulation using the impulsive approximation has been carried out and it predicts a displacement of the beam by  $\rm \Delta_{w}  = 17 - 33\,\mu m$. This is on the limit of our detector resolution and we will discuss ways of increasing $\rm \Delta_{w}$.  The simulation also indicates how we might observe the imaginary part of the weak value. 
\end{abstract}

\section{Introduction}

Weak (local) values were first considered by Landau \cite{landau} and London \cite{london} in connection with superfluids. As these values were not eigenvalues of the system and could not be measured in the usual way, they were not pursued. However Hirschfelder \cite{hirschfelder} subsequently realised the importance of local values, discussing them in terms of what he called ``subobservables". Moreover, Dirac \cite{dirac} had anticipated local values in his paper discussing non-commutative geometry in quantum mechanics.

The idea of a weak measurement has a long history \cite{holland} and was brought to prominence by Aharonov, Albert and Vaidman (AAV) \cite{yadalv88, yalv90} when they suggested a procedure to measure the weak value of spin for a particle experimentally. While ``local" was the original name for these variables and is still used by some authors \cite{berry}, ``weak" has become popular in this context. Although we believe ``weak" is misleading and can cause confusion with the electroweak force, we will continue to use it in the rest of this paper. 

Weak values are complex, in contrast to eigenvalues that  are only real. It must be clearly stated that the real part of the weak value is not to be identified as an eigenvalue. The experiment described here will show how both the real and imaginary parts of the weak value of spin may be observed.  Weak measurement can therefore reveal the more subtle details of quantum processes.

Measuring an eigenvalue uses a von Neumann (strong) measurement \cite{vn55}. This is a single stage process whereby the wave function is said to ``collapse".  In contrast, the weak measurement process has three stages; pre-selection, the weak stage and finally a strong stage (post-selection). 

The real parts of the weak values for the polarisation and momentum of photons \cite{nrrh91, skms11, skbb2} have been observed and measured. It should be noted that the theory of weak measurement was originally cast in the non-relativistic regime using \Schrs equation  (\Schr particles), whereas photons obey Maxwell's equations and are relativistic. In contrast to the photon case, the real and imaginary parts of the weak value of spin for non-relativistic neutrons have also been measured \cite{sponar}; we intend to  do the same for non-relativistic helium atoms. We are following a scheme outlined in AAV and by Duck, Stevenson and Sudarshan \cite{idpsgs89} which is a variant on the original Stern-Gerlach (S-G) apparatus \cite{wgos22}. A simulation has been carried out giving firm predictions of what should be observed within the scope of the parameters set by our experiment.

\section{Method for the weak measurement of spin for atomic systems} 

\subsection{Weak measurement of spin overview}

The weak measurement process allows for the detection of very small phase shifts. By preparing the system in a particular pre- and post-selected quantum state, it is possible to amplify these phase shifts so that they are more easily measurable. From this amplified signal, it is possible to abstract the desired observable of interest. As a consequence of this effect, the phrase ``weak value amplification" is commonly used in the literature. This amplification is constrained by certain conditions and owing to the limits imposed, these amplified shifts are still relatively small. 

\begin{figure}[H]
 \centering
  \includegraphics[width=5 in]{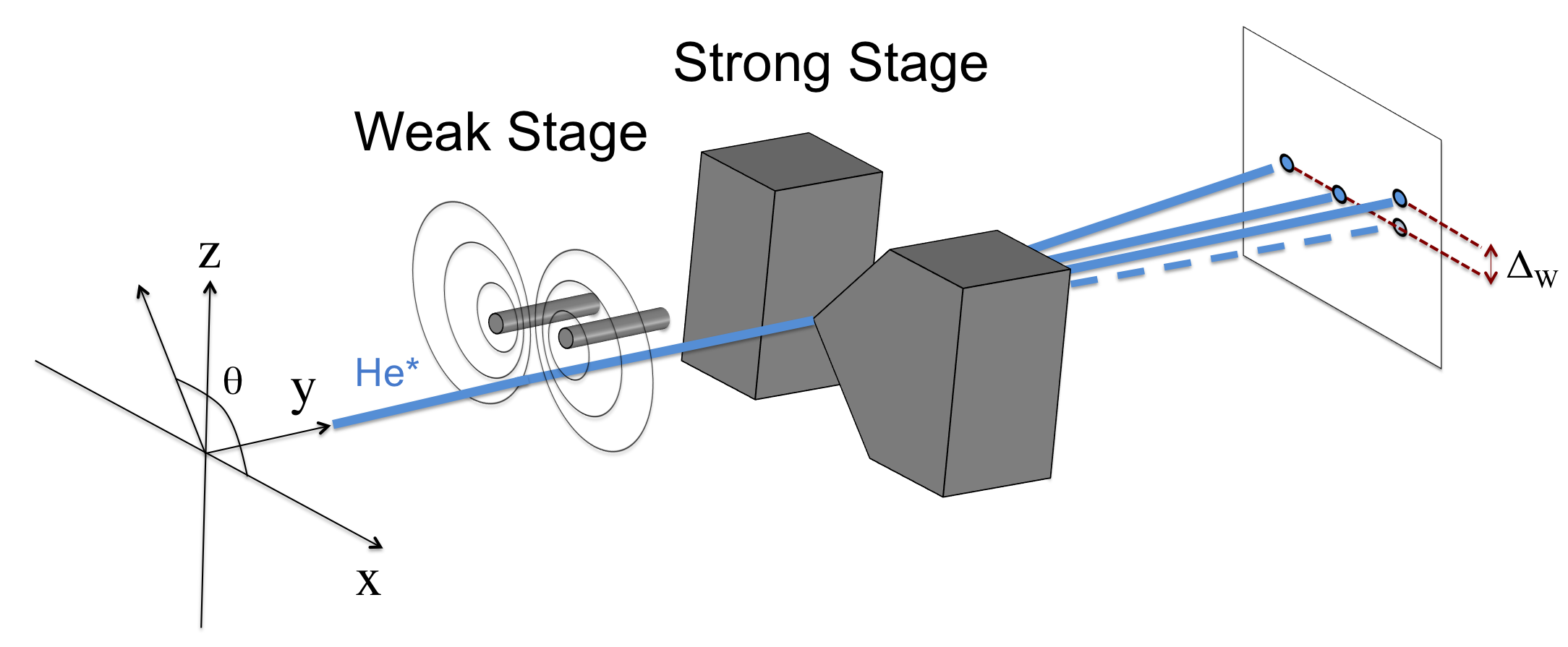}\caption{Schematic of the experimental technique. Helium atoms in the $\rm m = +1$ spin-up state enter from the left, with spin vector angle $\rm \theta$. The atoms pass through the weak and strong S-G magnets before eventually being incident on the detector. The displacement due to the weak measurement process is $\rm \Delta_w$. For simplicity the azimuthal angle $\rm \phi$ in the x-y plane is not shown.}  \label{fig:Setup-of-themyex}.
\end{figure}

The three stages of the weak measurement regime for spin are as follows. Atoms are first pre-selected in a desired spin state. In our case, we chose spin up at an angle $\theta$ in the x-z plane (see Fig.\ref{fig:Setup-of-themyex}). The atoms then propagate through the weak stage, comprising an S-G magnet with an inhomogeneous magnetic field in the z-axis. Here the field gradient is small, producing a small rotation of the spin vector. 

The strong stage consists of a second S-G magnet, with its inhomogeneous magnetic field aligned along the x-axis (see Fig.\ref{fig:Setup-of-themyex}). This field is large enough to separate the spin eigenstates on this axis. It is this separation that enables us to detect the small phase shift, proportional to $\rm \Delta_w$, induced by the weak stage as shown in the figure. The size of $\rm \Delta_w$ depends on various features of the apparatus. Furthermore, since this new shift is still relatively small, we must maximise it by suitably adjusting the experimental parameters as will be investigated below.

\subsection{Experimental realisation}

We have chosen to work with helium, excited into a metastable $\rm 2^{3}S_{1}$ triplet state. This form of helium has several advantages. 1. Its magnetic dipole moment has a magnitude of two Bohr magnetons \cite{baldwin, Halfmann}. This maximises the displacements produced by the S-G magnets. 2. It has a lifetime of approximately 8000\,s \cite{lifetimeHe}, being unable to decay via electric dipole transitions and the Pauli exclusion principle (i.e., its decay is doubly forbidden). This half-life provides sufficient time for the atoms to pass through all the stages of the apparatus before decaying. 3. Metastable helium atoms have an internal energy of 19.6\,eV, the highest of any metastable noble gas species. Upon collision with any surface, it will ionise with ease, allowing for detection with charged particle detectors. All of these characteristics will enhance the overall signal strength and sensitivity of the experiment.

\begin{figure}[H]
   \centering
   \includegraphics[width=5.5 in]{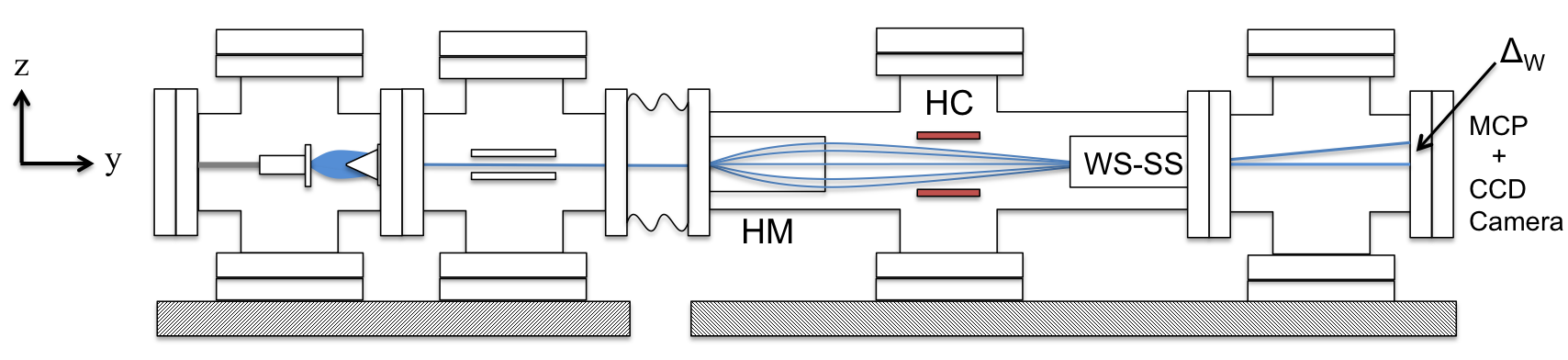} 
   \caption{The pulsed helium gas enters from the left. Preparation of the metastable atoms occurs on the first bench and the weak measuring process on the second. HM = hexapole magnet, HC = Helmholtz coils, WS = weak stage magnet, SS = strong stage magnet and MCP = micro-channel plate detector.}
   \label{fig: design_vincenzo_jan2016}
\end{figure}

Our experimental arrangement is shown in Fig. \ref{fig: design_vincenzo_jan2016}. Helium gas at high pressure enters the apparatus from the left and is pulsed using an electromagnetic valve producing a supersonic beam. The atomic beam is excited via an electron seeded discharge, where the atoms collide with a stream of energetic electrons in a 300\,V/cm electric field \cite{Halfmann}. The excited gas then passes through a $\rm 2\,mm$ diameter skimmer and travels between two electrically charged plates to remove residual ionised atoms and free electrons. A hexapole magnet focuses the $\rm m = +1$ state to a point along the axis of propagation, defocusing the $\rm m = -1$ state. The $\rm m = 0$ state is left untouched. The $\rm m = +1$  spin state is focused onto a second skimmer, producing an atomic beam with width $\rm 0.5-1\,\mu m$. 
 
 From this point onwards the apparatus is enclosed in a mu-metal shield to nullify the effects of the Earth's magnetic field. Using Helmholtz coils, the magnetic spin axis of the atoms is set at the angle $\rm \theta$ before entering the weak stage; the homogeneous magnetic field is aligned perpendicular to the atoms' direction of travel. Upon exiting of the strong stage, the atomic beam propagates freely onto a detector that consists of two micro-channel plates in a chevron configuration, coupled to a phosphor screen and CCD camera. The measured deflection $\rm \Delta_w$ will be proportional to the weak value of the atomic spin. 
 
Since the displacement  $\rm \Delta_w$ is very small, it is important to reduce vibrations in the final stages of the apparatus. To this end, two optical benches are used which are isolated by edge welded bellows; the first bench utilises turbo pumps in order to maintain adequate vacuum in the source chamber. The second, containing the preparation and detection chambers, employs zero vibration ion pumps.

\section{Simulation using the impulsive approximation} 

\subsection{Simulation}

The simulation is divided into three parts; the initial conditions, the application of the interaction Hamiltonian using the impulsive approximation \cite{bohm}, and the free evolution of the post-selected spin state. This approximation neglects the free evolution of the atoms in the weak magnet, only the interaction Hamiltonian is considered. It is also important to note that the inhomogeneous magnetic field produced by the S-G magnet in the weak stage is maximal along the z-axis, but negligible along the other two axes. The analysis follows the scheme outlined in \cite{idpsgs89}, but in our case we are using the spin-1, $\rm 2^{3}S_{1}$, metastable form of helium.  The helium atoms contain the $\rm m = 0$ spin state, but this is unaffected by inhomogeneous magnetic fields, passing through the experiment unhindered. 

\subsection{Initial conditions}

The helium gas is initially prepared as a pulsed beam and is described by the normalised Gaussian wave packet at time $t = 0$

\begin{equation}
 \psi(z, 0)=\frac{1}{\left(2\pi\sigma^{2}\right)^{\frac{1}{4}}}\exp\left({-\frac{z^{2}}{4\sigma^{2}}}\right),
 \label{eq:gaussian}
\end{equation}
where $\sigma$ is the width in position space.

Recall, before the atomic beam is sent through the weak stage, it is spin selected via a hexapole magnet. Shortly after, a pair of Helmholtz coils are used to set the pre-selected angle of spin $\theta$ within the x-z plane. We describe the resulting spinor in terms of polar angles $\theta$, and $\phi$, in the following form  \cite{Ballentine}

\begin{equation}
\bm{\xi}_{\rm i}(\theta ,\phi , 0)=\left[\begin{array}{c}
\frac{1}{2}(1+\sin(\theta))e^{-i\phi}\\
\frac{1}{\sqrt{2}}\cos(\theta)\\
\frac{1}{2}(1-\sin(\theta))e^{i\phi}
\end{array}\right]=\left[\begin{array}{c}
c_{\rm +}\\
c_{\rm0}\\
c_{\rm-}
\end{array}\right].\label{eq:si}
\end{equation}

The initial orientation of the spin vector angle $\theta$ can be seen in Fig.\ref{fig:Setup-of-themyex}, where the azimuthal angle $\phi$ (not shown), is the corresponding angle in the x-y plane. Angles $\rm \theta$ and $\phi$ can be changed by rotating the Helmholtz coils about the y and x-axes respectively. Therefore the initial wave function prior to entering the weak stage is

\begin{equation}
\Psi_{\rm i} (z, 0) = \psi(z, 0)\bm{\xi}_{\rm i}(\theta ,\phi , 0).
\end{equation}

The width of the atomic beam is set by passing it through an orifice/skimmer at the entrance of the weak stage.

\subsection{Simulation of the weak stage process}

The atoms then traverse the weak stage magnet, where the wave function evolves under the interaction Hamiltonian, weakly coupling the spin and centre-of-mass wave functions. The interaction Hamiltonian is given by

\begin{equation}
    H_{I}=\mu(\boldsymbol{\hat{s}}.\boldsymbol{B}),
\end{equation}

where $\hat{\boldsymbol{\bf s}}$ are the spin-1 matrices $\boldsymbol{\hat{s}}=[\hat{s}_{x},\hat{s}_{y}, \hat{s}_{z}]$, and the magnetic field  $\boldsymbol{B}=[B_{x},B_{y},B_{z}]$.  The inhomogeneous field in the z-direction is maximal $B_{z}=B_{0}+\frac{\partial B}{\partial z}z$, where $B_{0}$ is the homogeneous component of the magnetic field, and the fields in the other two directions can be neglected. Explicitly the interaction Hamiltonian is then

\begin{equation}
H_{I}=\mu\left(\begin{array}{ccc}
B_{z} & 0 & 0\\
0 & 0 & 0\\
0 & 0 & -B_{z}
\end{array}\right).
\end{equation}

At this point \Schrs equation is used to calculate the state of the system at a later time $\Delta t$, which is the time that the atom spends in the weak field. The resultant wave function is now given by 

\begin{equation}
\Psi_{{\rm w}}(z,\Delta t)=\exp\left(-\frac{i}{\hbar}\int_{0}^{\Delta t}H_{I}dt\right)\psi(z,0)\bm{\xi}_{\rm i}(\theta ,\phi , 0).
\end{equation}

Following the process of the weak measurement regime as described in \cite{Pan}, the pre-selected wave function is then post-selected via the strong stage into the spin-up, $\rm m = +1$ state in the x-basis $\bm{\xi}_{\rm f}^\dag(\pi ,0 , \Delta t)=\left[\begin{array}{ccc}\frac{1}{2} & \frac{1}{\sqrt{2}} & \frac{1}{2}\end{array}\right]$. Giving the final wave function 

\begin{equation}
\Psi_{{\rm f}}(z,\Delta t)=\bm{\xi}_{\rm f}^\dag(\pi ,0 , \Delta t)\exp\left(-i\frac{\mu\Delta t B_{z}\hat{s}_{z}}{\hbar}\right)\psi(z,0)\bm{\xi}_{\rm i}(\theta ,\phi , 0), 
\label{eq:phasebefore}
\end{equation}

explicitly this is

\begin{equation}
\Psi_{\rm f}(z,\Delta t)  = \psi(z,0)\Bigg[\frac{1}{2}\exp\Bigg(-i\frac{\mu\Delta t B_{z}}{\hbar}\Bigg)c_{+}+\frac{1}{\sqrt{2}}c_{0}+\frac{1}{2}\exp\Bigg(i\frac{\mu\Delta t B_{z}}{\hbar}\Bigg)c_{-}\Bigg].
\label{finalform}
\end{equation}

\subsection{Obtaining the weak value of spin}

The exponential (phase shift) in Eq. \ref{eq:phasebefore} can be Taylor expanded

\begin{equation}
\Psi_{{\rm f}}(z,\Delta t)=\bra{S_{{\rm f}}}\left[1-i\frac{\mu\Delta t B_{z}\hat{s}_{z}}{\hbar} -\frac{1}{2}\left(\frac{\mu\Delta t B_{z}\hat{s}_{z}}{\hbar}\right)^2+...\right]\ket{S_{{\rm i}}}\psi(z,0),\label{eq:allorder}
\end{equation}
where for convenience we have written   $\ket{S_{{\rm i}}}$ for $\bm{\xi}_{\rm i}$ and  $\bra{S_{{\rm f}}}$ for $\bm{\xi}_{\rm f}^\dag$. Hence
\begin{equation}
\Psi_{{\rm f}}(z,\Delta t)=\left[\braket{S_{\rm f}}{S_{\rm i}}-i\frac{\mu\Delta t B_{z} }{\hbar}\bra{S_{{\rm f}}}\hat{s}_{z}\ket{S_{{\rm i}}}-\frac{1}{2}\left(\frac{\mu\Delta t B_{z}}{\hbar}\right)^2\bra{S_{{\rm f}}}\hat{s}_{z}^2\ket{S_{{\rm i}}}+...\right]\\\psi(z,0)\label{eq:sfsi}.
\end{equation}

If the phase shift in Eq. \ref{eq:phasebefore} is sufficiently small such that the inequalities

\begin{equation}
\abs{\left(\frac{\mu\Delta t B_{z} }{\hbar}\right)^n\bra{S_{{\rm f}}}\hat{s}_{z}^n\ket{S_{{\rm i}}}} << \abs{\braket{S_{\rm f}}{S_{\rm i}}} \label{eq:limduckpan1}
\end{equation}

and

\begin{equation}
\abs{\left(\frac{\mu\Delta t B_{z} }{\hbar}\right)^n\bra{S_{{\rm f}}}\hat{s}_{z}^n\ket{S_{{\rm i}}} }<< \abs{\left(\frac{\mu\Delta t B_{z} }{\hbar}\right)\bra{S_{{\rm f}}}\hat{s}_{z}\ket{S_{{\rm i}}} }\label{eq:limduckpan1}
\end{equation}

hold true for $n\geq2$  \cite{idpsgs89,Pan}, then Eq. \ref{eq:sfsi} can be expanded to first order

\begin{equation}
\Psi_{{\rm f}}(z,\Delta t)=\left(\braket{S_{\rm f}}{S_{\rm i}}-i\frac{\mu\Delta t B_{z} }{\hbar}\bra{S_{{\rm f}}}\hat{s}_{z}\ket{S_{{\rm i}}}\right)\\\psi(z,0)\label{eq:oneorder},
\end{equation}

and the transition probability amplitude $\braket{S_{\rm f}}{S_{\rm i}}$ factored out

\begin{equation}
\Psi_{{\rm f}}(z,\Delta t)=\braket{S_{\rm f}}{S_{\rm i}}\left(1-i\frac{\mu\Delta t B_{z} }{\hbar}\frac{\bra{S_{{\rm f}}}\hat{s}_{z}\ket{S_{{\rm i}}}}{\braket{S_{\rm f}}{S_{\rm i}}}\right)\\\psi(z,0)\label{eq:factorout}.
\end{equation}

If the inequality 

\begin{equation}
L=\abs{\left(\frac{\mu\Delta t B_{z} }{\hbar}\right)\frac{\bra{S_{{\rm f}}}\hat{s}_{z}\ket{S_{{\rm i}}} }{\bra{S_{\rm f}}\left.S_{\rm i}\right\rangle}}<< 1
\label{eq:limpanpaper}
\end{equation}

is also true \cite{idpsgs89,Pan}, where $L$ is a limit to be determined, then Eq. \ref{eq:factorout} can be cast back into exponential form

\begin{equation}
\Psi_{\rm f}(z,\Delta t)=\bra{S_{\rm f}}\left.S_{\rm i}\right\rangle \psi(z,0)\exp\left(-i\frac{\mu\Delta t B_{z}}{\hbar}W\right)\label{eq:finalweaknofree},
\end{equation}

where the weak value is $W=\frac{\bra{S_{{\rm f}}}\hat{s}_{z}\ket{S_{{\rm i}}} }{\bra{S_{\rm f}}\left.S_{\rm i}\right\rangle}$. 

Using the defined pre- and post-selected states, the transition probability amplitude $\braket{S_{\rm f}}{S_{\rm i}}=\cos(\phi)+\cos(\theta)-i\sin(\phi)\sin(\theta)$ and therefore the final post-selected wave function becomes

\begin{equation}
\Psi_{\rm f}(z,\Delta t)=\bra{S_{\rm f}}\left.S_{\rm i}\right\rangle \psi(z,0)\exp\left[-i\frac{\mu\Delta t B_{z}}{\hbar}\left(\frac{\sin(\theta)\cos(\phi)-i\sin(\phi)}{\cos(\phi)+\cos(\theta)-i\sin(\phi)\sin(\theta)}\right)\right].\label{eq:fullweakexp}
\end{equation}

It can be seen by comparing Eq. \ref{eq:fullweakexp} and Eq. \ref{eq:phasebefore} that the phase shift is now proportional to the weak value of spin, and this phase shift can be varied by changing the spin vector angles $\theta$ and $\phi$. Note $W$ is a complex number whose real and imaginary parts are

\begin{equation}
W_{Re}=\frac{\sin(\theta)}{1+\cos(\phi)\cos(\theta)},
\end{equation}
and

\begin{equation}
W_{Im}=-\frac{\sin(\phi)\cos(\theta)}{1+\cos(\phi)\cos(\theta)}.
\end{equation}

\begin{figure}[H]
   \centering
   \includegraphics[width=5 in]{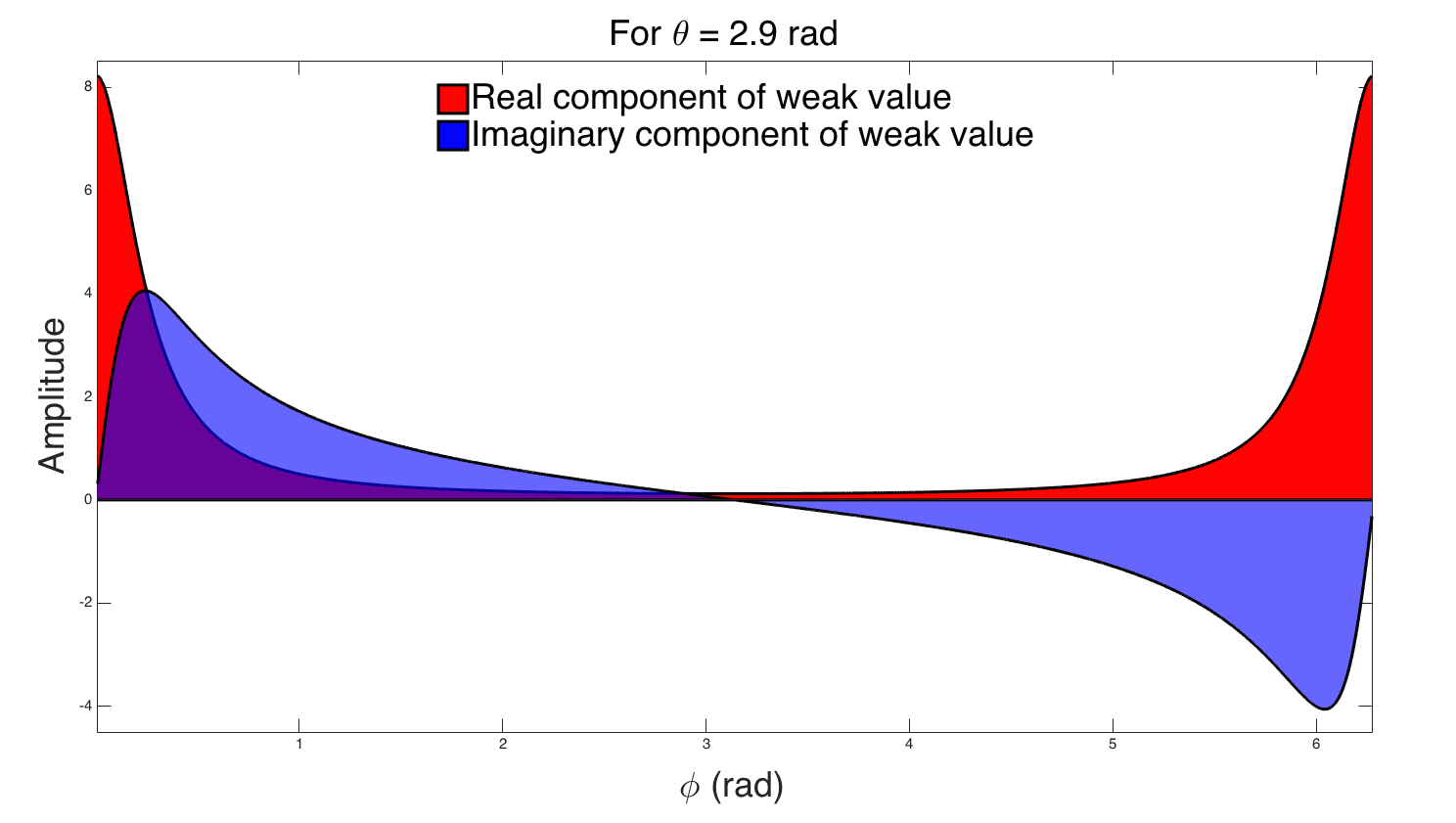} 
   \caption{Plot showing the real (red) and the imaginary (blue) components of the weak value of spin for a fixed angle $\theta= \rm 2.9\,rad$ while varying the azimuthal angle $\rm \phi$.}
   \label{fig: realIm}
\end{figure}

It should also be noted that if $\phi = 0$, then the real part reduces to $\tan\left(\frac{\theta}{2}\right)$ and the imaginary part is zero. In this experiment, the real part of the weak value of spin will be measured by setting $\phi = 0$ and varying the angle $\theta$ between 0 and 2$\pi$. The imaginary part can then be observed by fixing the angle $\theta$ and allowing the angle $\phi$ to vary. Plots of the two functions are shown in Fig. \ref{fig: realIm} for a fixed angle $\theta = \rm 2.9\,rad$.

\subsection{Free evolution of the Gaussian wave packet at the detector}

After the strong stage, the problem is treated as the free evolution of a Gaussian wave packet by solving the Pauli equation using well-known methods \cite{bohm, dewdspin}. Polar decomposing the wave function 

\begin{equation}
\Psi(z,t)=R(z,t)\exp\left(i\frac{\Phi(z,t)}{\hbar}\right),
\end{equation}

we find

\begin{equation}
R(z,t)=\bra{S_{f}}\left.S_{i}\right\rangle \left[2\pi\sigma^{2}\left(1+\frac{\hbar^{2}t^{2}}{4m^{2}\sigma^{4}}\right)\right]^{-\frac{1}{4}}\exp\left[\frac{-\left(z+utW_{Re}\right)^{2}}{4\sigma^{2}\left(1+\frac{\hbar^{2}t^{2}}{4m^{2}\sigma^{4}}\right)}+W_{Im}\frac{\mu\Delta t}{\hbar}B_{z}\right]
\end{equation}

\begin{small}
\begin{equation}
\frac{\Phi(z,t)}{\hbar}=-\frac{\mu\Delta t}{\hbar}\left[B_{0}W_{Re}+\frac{\partial B}{\partial z}\left(W_{Re}z+\frac{1}{2}utW_{Re}^{2}\right)\right]-\frac{1}{2}\arctan\left(\frac{\hbar t}{2m\sigma^{2}}\right)+\frac{\hbar t\left(z+utW_{Re}^{2}\right)}{8m\sigma^{4}\left(1+\frac{\hbar^{2}t^{2}}{4m^{2}\sigma^{4}}\right)}.
\end{equation}
\end{small}

It can be seen that if $\phi=0$, then $W_{Im}=0$ and the mean of the post-selected wave function shifts by the value $\Delta_{\rm w} = \big(\frac{\mu}{m}\frac{\partial B}{\partial z}\Delta t\big)t\tan\left(\frac{\theta}{2}\right) = utW_{Re}$, where $u$ is the transverse velocity of the helium atoms. This is in contrast to the standard S-G experiment where the shift is only $ut$.  

The probability density can now be computed, giving the form of the wave function as seen by the detector
\begin{equation}
\begin{small}
|\Psi_{\rm D}(z,t)|^{2}=|\bra{S_{\rm f}}\left.S_{\rm i}\right\rangle |^{2}\left[2\pi\sigma^{2}\left(1+\frac{\hbar^{2}t^{2}}{4m^{2}\sigma^{4}}\right)\right]^{-\frac{1}{2}}\exp\left[\frac{-\left(z+utW_{Re}\right)^{2}}{2\sigma^{2}\left(1+\frac{\hbar^{2}t^{2}}{4m^{2}\sigma^{4}}\right)}+2W_{Im}\frac{\mu\Delta t}{\hbar}B_{z}\right]. \label{eq:waveprob}
\end{small}
\end{equation}

As the pre- and post-selected spin states approach orthogonality, $\theta$ tends to $\pi$, $\Delta_{\rm w}$ increases but the transition probability decreases. This reduces the number of post-selected events of interest, leading to the need for longer experimental runs.

Again it is important to understand that this effect only arises when the phase shift acquired at the first stage is sufficiently small (see Eq. \ref{eq:limpanpaper}). The centre-of-mass wave function is displaced but its overall shape is maintained after exiting the weak stage. 

\subsection{The limit and its validity for the real part of the weak value of spin, $\boldsymbol{\rm\phi=0}$}

The definition of the real part of the weak value used in the literature, $\tan(\frac{\theta}{2})$, only considers the first order Taylor expansion of the phase shift acquired at the weak stage, when $\phi = 0$. This approximation is sufficient when describing an ideal experiment, but in reality $L$ (given by Eq. \ref{eq:limpanpaper}) is constrained by the apparatus variables,  such as the beam width, $\sigma$, before the atoms enter the weak stage and the time that the atoms are in the weak stage, $\Delta t$. If $L$ exceeds a certain threshold, then the approximation breaks down as the higher order terms dominate. The new limit is 

\begin{equation}
{L} = \frac{\mu\Delta t \left(B_{0}+\frac{\partial B}{\partial z}\sigma \right)}{\hbar}\tan\left(\frac{\theta}{2}\right)  <<1 \label{eq:lim},
\end{equation}

where $z$ is now directly related to the spread of the beam, $\sigma$, in the inhomogeneous magnetic field \cite{idpsgs89,Pan}. 

In order to describe the experimental data effectively (in terms of the first order approximation), it is important to analyse the limit at the point that the approximation breaks down in detail. This maximum limit can be determined by analysing the weak measurement process for two Gaussian wave packets, one describing the first order approximation (Eq. \ref{eq:waveprob}), and the other, an exact case where no approximation is considered, is calculated via the time evolution of Eq. \ref{eq:phasebefore}.

\begin{figure}[H]
   \centering
   \includegraphics[width=6 in]{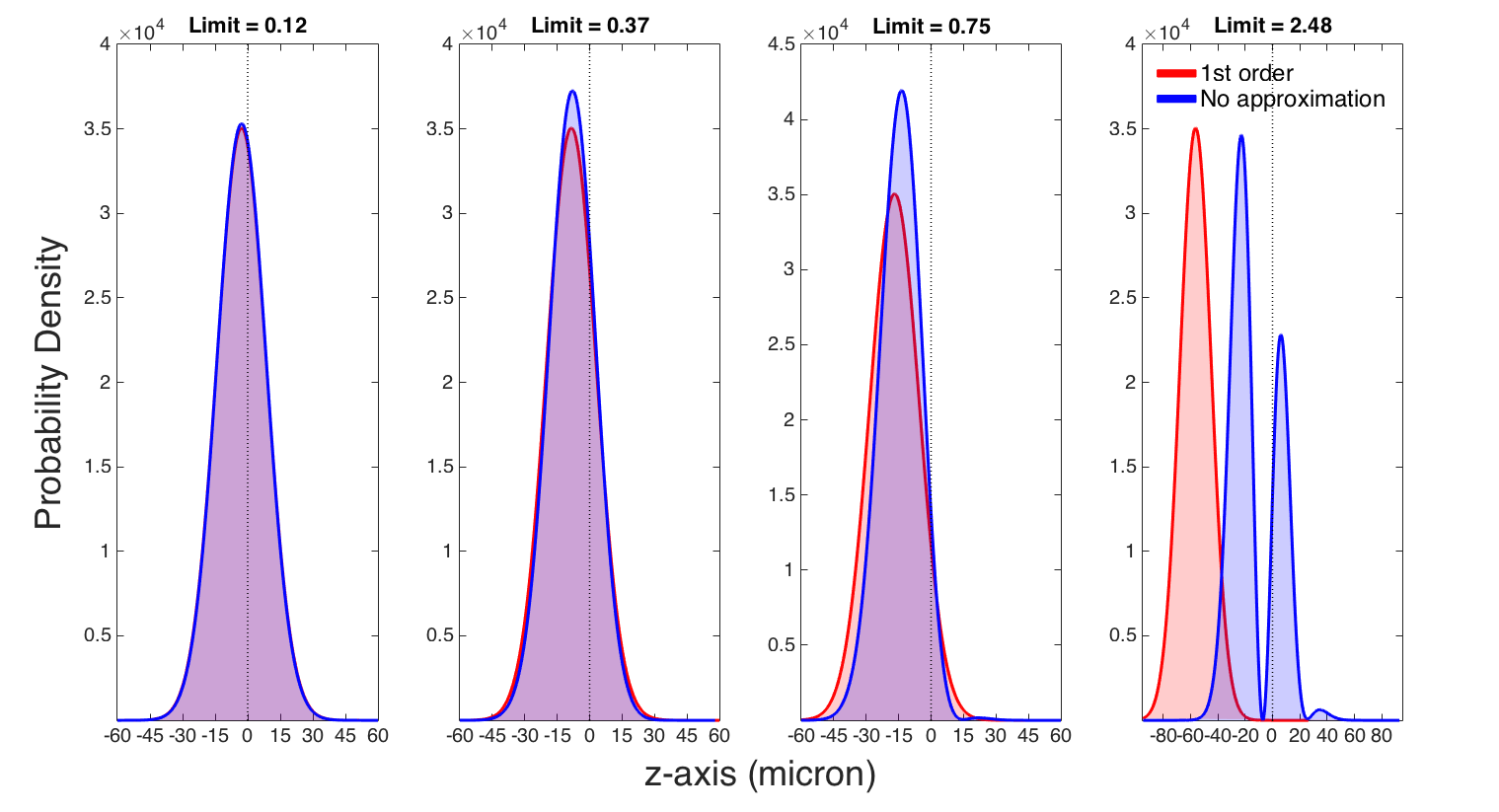} 
   \caption{A series of plots showing how the displacement, $\Delta_{\rm w}$, of the Gaussian wave packet is constrained by various limits. The red curve is the first order approximation which is dominated by $\tan(\frac{\theta}{2})$. The blue curve is the exact treatment of the system taking into account the higher order terms. The red and blue curves coincide when $L = 0.37$; this is the maximum limit for which the first order approximation holds.}
   \label{fig: Limit}
\end{figure}

The first order approximation cut-off is calculated by finding the last point at which the two simulations coincide; past this point, the simulations begin to deviate from one another due to the higher order effects. Fig. \ref{fig: Limit} shows both simulations for a selection of limits; the limits are calculated by increasing the inhomogeneous magnetic field in the weak stage only, all other variables are held constant. The beam width is set to 1 $\mu$m before the weak stage, and the weak stage itself is 10 mm in length. The distance from the weak magnet's exit to the detector is 2.5 m, and the expected atomic velocity of 1750 m/s was used.
  
  From the statistics, the desired limit is 0.37. Past this, the two Gaussian wave packets deviate dramatically. The mean of the first order approximation, red curve, continues to increase, while the blue curve, no approximation, comes to a halt and slowly reverts to that of a standard S-G measurement. The limit of 0.37 and Eq. \ref{eq:lim} allow for the optimisation of the experiment and permit the experimental parameters to be set in order to attain the largest possible deviation. This also enables the experiment to be described in terms of the well-known first order approximation.

With the maximum limit calculated, it is possible to analyse the system in a way that does not involve individual variables, such as the inhomogeneous field strength of the weak stage, or the time that the atom spends in this field. By setting Eq. \ref{eq:lim} to the desired limit and neglecting the homogeneous component of the magnetic field 

\begin{equation}
{L}=\frac{\mu\Delta t\frac{\partial B}{\partial z}\sigma}{\hbar}\tan\left(\frac{\theta}{2}\right)=0.37.
\label{limit037}
\end{equation}

The displacement of the Gaussian wave packet can then be calculated
in terms of $L$. This is given by

\begin{equation}
\Delta_{{\rm w}}=\frac{\mu\frac{\partial B}{\partial z}(\Delta t)t}{m}\tan\left(\frac{\theta}{2}\right)=\frac{\hbar t}{\sigma m}{L}\label{eq:fixedl}.
\end{equation}

In order to maximise the displacement, the flight time of the atoms after the weak stage, $t$, must be as large as possible, while the width of the beam, $\sigma$, before the weak stage must be as small as possible. The correct combination of these two variables will ultimately enable this small displacement to be more easily resolved. 

From Eq. \ref{eq:fixedl} it can be seen that by fixing $\sigma$ and $t$, the displacement of the beam is determined solely by the limit imposed from the previously discussed simulation. $\sigma$ will be constrained by the smallest orifice presently available, and $t$ is limited by the size of the lab and the velocity of the metastable helium atoms. By adjusting the spin vector angle $\theta$, but maintaining $L$ = 0.37 to produce maximum displacement,  $\Delta_{\rm w}$  can be measured against  varying $\theta$. And from this, the real part of the weak value for the first order approximation, $\tan\left(\frac{\theta}{2}\right)$, can be obtained. Note that as long as $L$ is kept constant, regardless of the change in $\theta$, the displacement $\Delta_{{\rm w}}$ will also always be constant if both $t$ and $\sigma$ are unchanged.

Substituting Eq. \ref{eq:fixedl} into Eq. \ref{eq:waveprob}, we can see how the system behaves with this set limit

\begin{equation}
|\Psi_{\rm D}(z,t)|^{2}=|\bra{S_{\rm f}}\left.S_{\rm i}\right\rangle |^{2}\left[2\pi\sigma^{2}\left(1+\frac{\hbar^{2}t^{2}}{4m^{2}\sigma^{4}}\right)\right]^{-\frac{1}{2}}\exp\left[\frac{-\left(z+\frac{\hbar t}{\sigma m}{L}\right)^{2}}{2\sigma^{2}\left(1+\frac{\hbar^{2}t^{2}}{4m^{2}\sigma^{4}}\right)}\right]. \label{eq:waveprobnew}
\end{equation}

\begin{figure}[H]
   \centering
   \includegraphics[width=6 in]{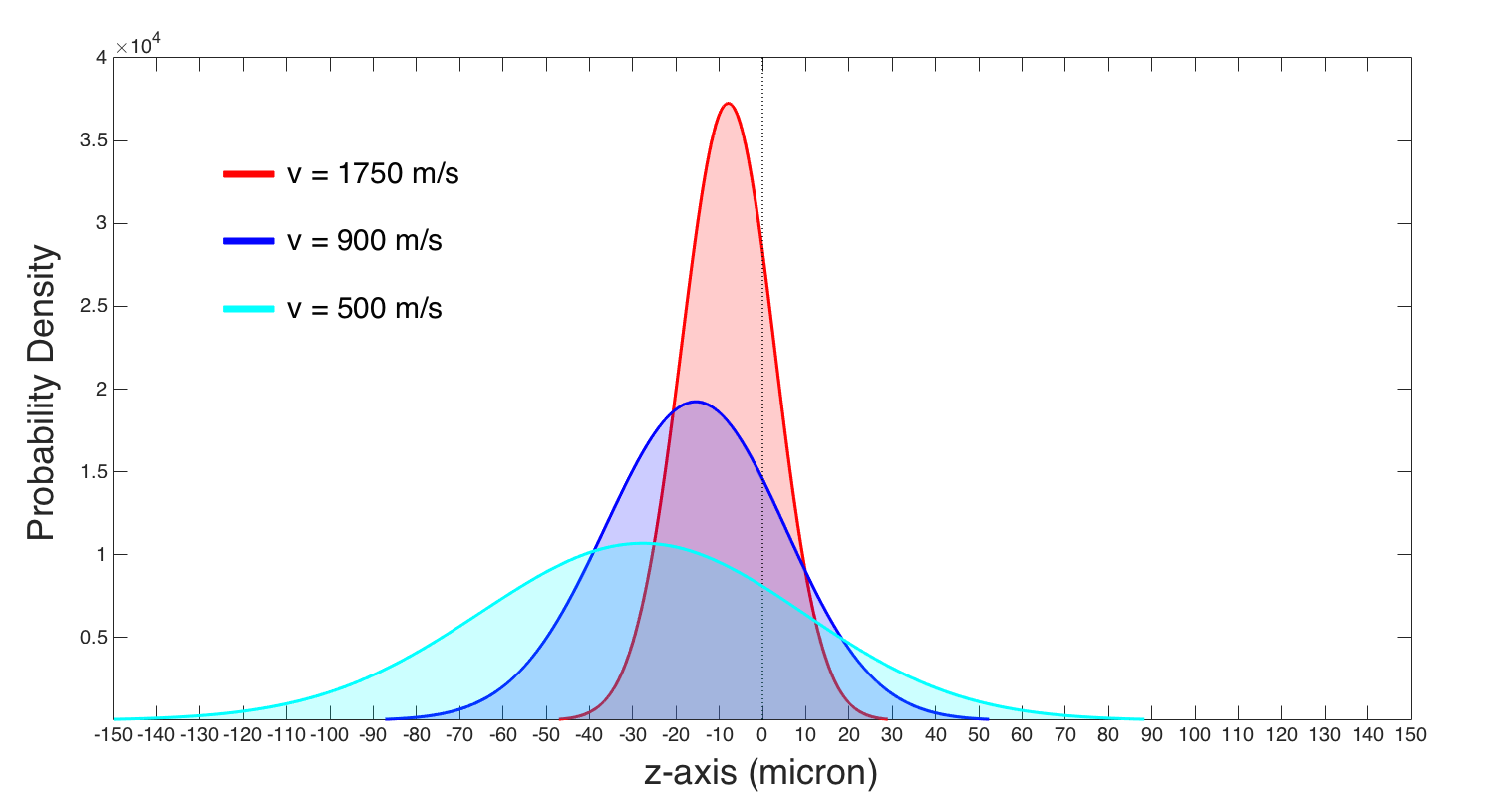} 
   \caption{Plot showing the displacement of the Gaussian wave packet due to the weak measurement process for three different atomic velocities at the limit of $L= 0.37$.}
   \label{fig: velocityplot}
\end{figure}

By again setting the orifice size to 1 $\mu$m, and the distance from the weak magnet exit to the detector to 2.5 m, Eq. \ref{eq:waveprobnew} can be plotted at a fixed limit of 0.37 for varying atomic velocities. It can be shown by looking at Fig. \ref{fig: velocityplot}, that the time taken for the atoms to reach the detector may enhance $\Delta_{\rm w}$ without compromising the limit on the higher order terms. 

\begin{figure}[H]
   \centering
   \includegraphics[angle=270, width=3 in,angle =90]{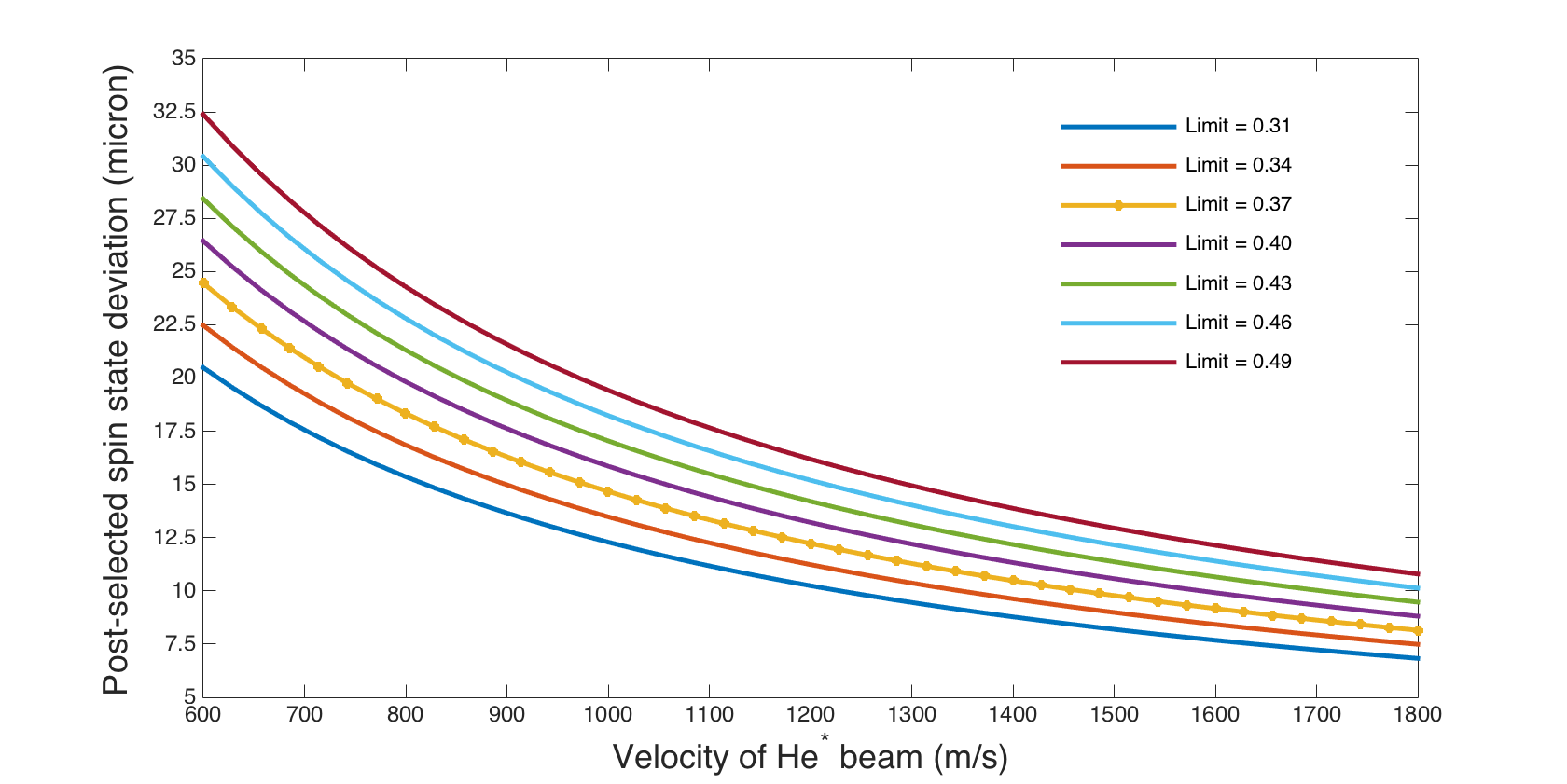} 
   \caption{Plot showing the displacement, $\Delta_{\rm w}$, as a function of  the velocity of the atoms with $L= 0.37$ highlighted.}\label{fig: velocitybeamdev}
\end{figure}

In order to increase the shift  $\Delta_{{\rm w}}$ further, the limit $L$ can be increased. Note, this introduces errors into the analysis as the first order approximation may no longer be valid.  
Fig. \ref{fig: velocitybeamdev}  illustrates how changing the limit marginally while decreasing the velocity, allows for larger measurable shifts.

Ideally one would want to stay as close to the calculated limit as possible, but reduce the atomic beam velocity in order to attain shifts that are more easily measurable on the selected detection device.

\subsection{Expected deviation with updated experimental parameters}

\begin{figure}[H]
   \centering
   \includegraphics[width=0.9\textwidth]{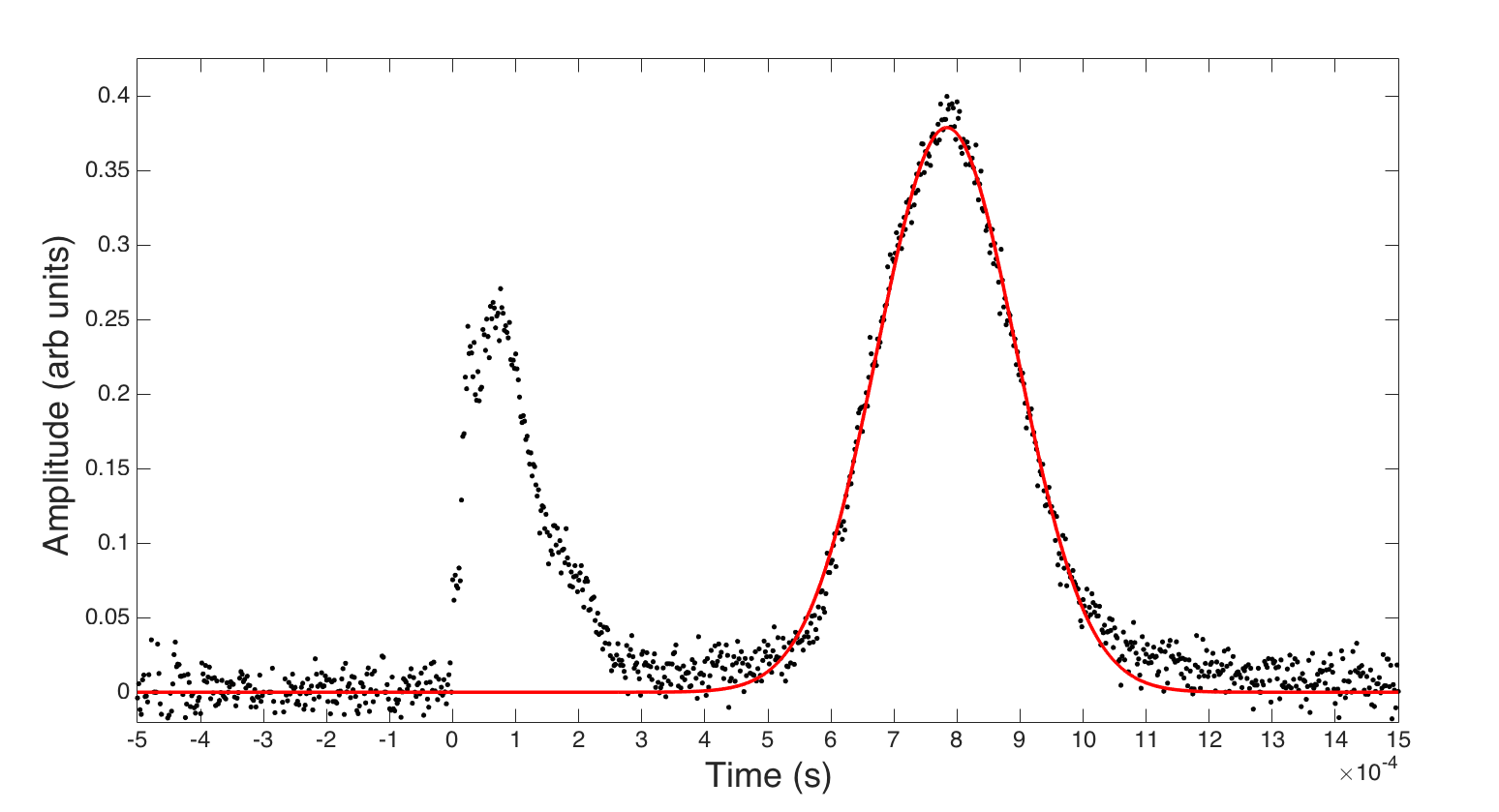} 
   \caption{TOF measurement from the MCP detector showing two visible peaks, the first due to photons from the pulsed discharge, and the second, metastable helium atoms arriving shortly after. From this data the mean velocity of the metastable helium atoms is calculated to be 1717\,m/s.}
   \label{fig:TOFplot}
\end{figure}

To obtain numerical predictions for the expected experimental deviation, updated parameters from preliminary experiments need to be considered. An important factor is the velocity of the atomic beam. This is obtained from time of flight (TOF) measurements (see Fig. \ref{fig:TOFplot}). Fig. \ref{fig:TOFplot} shows two distinct peaks. The first measured peak is the photon signal arriving from the pulsed discharge; this photon pulse is used to zero the timing signal as it arrives instantaneously as the valve is open. Shortly after the photon peak, metastable helium atoms are detected. By knowing the flight distance from the pulsed valve to the detector, in this case 1.35 m, and the time from the rising edge of the photon peak to the atomic signal, the velocity of the atoms can be calculated. From the measured data, the mean velocity of the metastable helium atoms is found to be approximately 1717 m/s. 

A list of the variables that will influence the expected atomic deviation is given in Table \ref{table:data}. Note, the beam width $\sigma$ has been reduced to 0.5 $\mu\rm m$ in order to achieve the greatest possible shift, while the distance from the weak stage exit to the detector is still 2.5 m as this is constrained by the size of the laboratory. 

\begin{table}[H]
\caption{List of variables used in the final experiment in order to measure the real part of the weak value of spin.}
\centering
\begin{tabular}{l | c c }
\hline\hline
  & Magnitude & units\\ [0.5ex]
\hline
Velocity of the metastable helium atoms &1717  & m/s\\
First order approximation limit & 0.37 \\
Distance of free flight from the weak stage exit to the MCP detector & 2.5 & m\\
Azimuthal angle $\phi$ & 0 & rad\\ 
Beam width just before the weak stage  $\sigma$ & 0.5 & $\mu\rm m$\\
Temperature of the pulsed gas  & 293 & K\\ [ 1ex]
\hline
\end{tabular}
\label{table:data}
\end{table}

Looking at Eq. \ref{eq:fixedl} and using the variables given in Table \ref{table:data}, it can be seen that the shift along the z-axis due to this process is $\rm \Delta_{w}  = 17\,\mu m$. As pointed out previously, reducing the velocity of the atomic beam increases the shift. By lowering the temperature of the nozzle by means of a liquid nitrogen cryostat, it is possible to reduce the velocity of the beam to approximately $\rm 1200 - 900\,m/s$. This decrease in velocity, while maintaining the chosen limit, would increase the shift to $\rm 24-33\,\mu m$.

\clearpage

\section{Conclusion}

The experiment described in this paper is designed to measure the real part of the weak value of spin. In order to achieve meaningful results, a full simulation of the process has been carried out. This includes the optimisation of the experimental parameters in order to achieve maximum resolution, which is necessary as the limits imposed by the theory cause the measured shift to be relatively small.  

Investigating this limit has determined the range over which the first order approximation holds. We have analysed and chosen the experimental parameters to achieve the largest possible displacement.  A full simulation has been carried out predicting a shift $\Delta_{\rm w}$, of between 17 and 33 \,$\rm \mu m$. The predicted upper bound displacement (33 \,$\rm \mu m$) is within the experimental resolution, with scope to increase this further by additional cooling of the atomic beam. It has also been shown how both the real and imaginary parts of the weak value can be observed by analysing their distributions in $\theta$ and $\rm \phi$. 

Our experiment is designed to confirm that the weak value of the spin measures the angle of polarisation of the original spin system. This will provide us with a tool for measuring the polarisation angles of spin systems in general.  For example, the spin orientation calculations of Dewdney et al. \cite{dehokyvi, dehoky} show that for an EPR-entangled pair, when one of the pair enters a magnetic field, its spin polarisation angle becomes established.  The other remote partner simultaneously becomes polarised in the opposite direction even though it is in a field-free region.  By measuring the weak value of the spin of this distant partner, we would, in principle, be able to demonstrate the non-local action of the quantum potential.

The authors would like to thank the Fetzer Franklin Fund of the John E. Fetzer Memorial Trust for their continued generous support.

\end{document}